\documentclass[aps,nofootinbib,longbibliography,prd,twocolumn]{revtex4-1}
\usepackage[babel]{csquotes}
\usepackage{graphicx}
\usepackage{amsmath,amssymb}
\usepackage[colorlinks,citecolor=blue,linkcolor=blue,urlcolor=blue]{hyperref}
\usepackage{mathrsfs}
\usepackage{enumerate}
\def\be{\begin{equation}}
\def\ee{\end{equation}}

\newcommand{\SU}{\mathsf{SU}}

\newcommand{\SL}{\mathsf{SL}}

\newcommand{\HH}{{\mathcal{H}}}

\DeclareMathOperator{\trace}{tr}

\newcommand*{\CC}{\mathbb{C}}

\usepackage[inline]{enumitem} 
\setlist[itemize]{noitemsep} 

\usepackage{todonotes}

\newcommand{\wtj}[6]{
\begin{pmatrix}
   #1 & #2 & #3 \\
   #4 & #5 & #6
\end{pmatrix}
}

\newcommand{\wfj}[9]{
\begin{pmatrix}
   #1 & #2 & #3 & #4 \\
   #5 & #6 & #7 & #8
\end{pmatrix}^{(#9)}
}
 
\newcommand{\wsj}[6]{ 
\begin{Bmatrix}
   \,\,\,#1 & #2 & #3\,\,\, \\[2pt]
   \,\,\,#4 & #5 & #6 \,\,\,
\end{Bmatrix}
}
 
\newcommand{\wnj}[9]{
\begin{Bmatrix}
   \,\,\,#1 & #2 & #3\,\,\, \\[2pt]
   \,\,\,#4 & #5 & #6\,\,\, \\[2pt]
   \,\,\,#7 & #8 & #9\,\,\,
\end{Bmatrix}
}

\begin{document}

\title{Primordial fluctuations from quantum gravity}

\author{Francesco Gozzini\,$^{a,b}$ \ and \ Francesca Vidotto\,$^{c,d}$ \vspace{2mm}}

\affiliation{\small\mbox{$^a$}
CPT, Aix-Marseille\,Universit\'e, Universit\'e\,de\,Toulon, CNRS, 13288 Marseille, France
\\
\small\mbox{$^b$}
Dipartimento di Fisica, Universit\'a di Trento, and TIFPA-INFN, 38123 Povo TN, Italy
\\
\small\mbox{$^c$}
Department\,of\,Applied\,Mathematics, University\,of\,Western\,Ontario, London, ON N6A\,5B7, Canada 
\\
\small\mbox{$^d$}
University\,of\,the\,Basque\,Country UPV/EHU, Departamento\,de\,F\'isica\,Te\'orica, 48940 Leioa, Spain }


\begin{abstract}
\noindent We study fluctuations and  correlations between spacial regions, generated by the primordial quantum gravitational phase of the universe. We do so  by a numerical evaluation of Lorentzian 
amplitudes in Loop Quantum Gravity, in a non-semiclassical regime.  We find that the expectation value of the quantum state of the geometry emerging from the early quantum phase of the universe is a homogeneous space but fluctuations are very large and correlations are strong, although not maximal.  In particular, this suggests that early quantum gravitational effects could be sufficient  to solve the cosmological horizon problem.  
\end{abstract}
\maketitle

\section{Introduction}

The universe emerged from its early quantum gravitational phase in a state that included fluctuations with  correlations between distinct regions of space. These primordial fluctuations play a key role in cosmology --- with or without inflation --- in particular as seeds for structure formation. Here we explore the physical genesis of these fluctuations from the primordial quantum gravitational cosmological state. We use Loop Quantum Gravity (LQG) and a simple model of the early universe. 

We consider the quantum transition from an empty state to a 3-geometry. As argued in \cite{Bianchi2010b,Borja2012a,Vidotto2017}, this transition may be relevant in a bounce cosmology, because the empty state can be interpreted as the dominant intermediate state in the transition through the bounce. Such transition may also describe the very early phase of a Big Bang cosmology, as the bridge from `nothing' to an initial quantum geometry. We treat the dynamics of the gravitational degrees of freedom non-perturbatively, using the covariant formulation of the dynamics of LQG. This is a fully Lorentzian calculation. This dynamics can describe the quantum tunneling from nothing to a 3-geometry.  The quantum state generated by this transition determines a probability distribution over 3-geometries, which includes correlations between spatial regions.  
We truncate the degrees of freedom of the gravitational field to a small finite number in addition to the scale factor. We view this as a finer truncation than the common symmetry reduction that treats only the scale factor non perturbatively \cite{Borja2012a,Vidotto2017}.  Even for this simple model, exact analytical results are hard, hence we use numerical methods. We report here the implementation of the calculations and three results.  

The first regards the expectation value of the variables describing the geometry at a given value of the scale factor. The result is that this turns out to be rather precisely given by (the truncation of) a metric 3-sphere. At first this result may seem pretty obvious, for reasons of symmetry, but combined with what follows it is much less so.

\pagebreak

The second result is that (contrary to our initial expectation), the variance of these variables is very large. This variance cannot be too small, because the variables do not commute, but one might have expected a coherent state minimizing the spread. Instead, the spread is almost maximal.  This means that (in spite of what the first result mentioned above might misleadingly suggest), the quantum state that emerges from the big bang in this approximation is \emph{not} peaked on a homogeneous geometry: fluctuations are ample.

Finally, we study the correlation between distinct spatial regions. We look directly at the correlation between two variables in distinct regions, as well as, more generally, at the entanglement entropy between one region and the rest of the geometry. We find that the correlation is strong and does not vanish with the growth of the scale factor. Moreover, the entanglement entropy has a highly non-trivial behaviour, apparently reaching asymptotically a stable value.

All this indicates that the universe emerging from an early quantum era has a rich structure of fluctuations, homogeneity properties, and large scale correlations, due to the common quantum origin of spatially separated regions, which can be studied theoretically and appear to be compatible with the observed universe. Inflation or a bounce, in particular, might not be strictly necessary to circumvent the horizon problem. That is, the result of classical general relativity stating that distant regions of the universe were causally disconnected in the past in a (non-inflationary) expansion is not anymore true if the initial quantum phase is taken into account. 

The numerical calculation we develop uses the amplitude of covariant Loop Quantum Gravity in a non-semiclassical regime, which can be loosely understood as a quantum `tunneling' transition.  We find that the relevant contribution from the loop transition amplitude comes from the so-called \emph{vector geometries} \cite{Barrett2009, Dona2019}. This clarifies their physical meaning as a non-suppressed component of the amplitude, that appeared quite mysterious so far. 

The model we analyse is crude and insufficient to develop a finer quantitative analysis. In particular, the truncation we use implies that any two regions are in contact, and this prevents us from distinguishing short from long distance correlation. We leave open an intriguing question, which might be addressed with these methods: could the quantum epoch directly yield almost scale-invariant cosmological fluctuations?


\section{Structure of the model}

\subsection{Truncation}

We discretize a closed cosmological 3-geometry into five tetrahedra glued to one another, giving an $S_3$ topology. This is the simplest regular triangulation of a topological 3-sphere, and it corresponds to the boundary of a geometrical 4-simplex. This geometry has twenty degrees of freedom, which capture the gravitational field in this truncation. To first order in the spinfoam expansion, the result of the transition from nothing to a 3-geometry is described in this truncation directly by the covariant LQG Lorentzian vertex amplitude.  

We take the further simplification consisting in assuming all areas of all the faces of the tetrahedra to be equal and we use this common value as a proxy for the scale factor. The remaining degrees of freedom are those characterizing the shapes of the five tetrahedra. We are interested in the fluctuations of these variables and the correlations between (variables in) distinct tetrahedra, at different values of the scale factor.


\subsection{Quantum theory}

In Loop Quantum Gravity, the state space of a truncation of the degrees of freedom of the gravitational field is given by the Hilbert space 
\begin{equation}
  \HH = L^2[\SU(2)^L / \SU(2)^N]_\Gamma
\end{equation}
where $\Gamma$ is an oriented graph with $L$ links and $N$ nodes that determines the way $\SU(2)^N$ acts on $\SU(2)^L$ \cite{Rovelli2015a}. Here we consider the truncation of general relativity defined by the complete graph $\Gamma$ with five nodes, hence $N=5$ and $L=10$. We label the nodes as $n=1,\ldots,5$. We label the (oriented) links as $l=1,\ldots,10$, or alternatively in term of the two nodes they link: $l=nn'$. Geometrically, this graph corresponds to the discretization of a three-sphere with five tetrahedra, joined via their 10 shared faces. 

The spin network basis in $\HH$ is given by the states 
\begin{equation}
  |j_l, i_n\rangle
\end{equation}
where the $j_l$'s are spins (half-integer values labelling $\SU(2)$ irreps) and the $i_n$'s are a basis in the corresponding intertwiner space
\begin{equation}
{\cal I}_n= \left({\bigotimes}_{n'\ne n}\ V_{j_{nn'}}\right) \Big/\ \SU(2).
\end{equation}
where $V_j$ is the spin-$j$ representation space of $\SU(2)$. We shall focus on the subspaces $\HH_j$ of $\HH$ defined by $j_l=j$. These have the tensorial structure
\begin{equation}
\label{eq:j-subspaces}
\HH_j={\bigotimes}_n \ {\cal I}_n
\end{equation}
where each ${\cal I}_n$ is isomorphic to $(V_j {\otimes}V_{j} {\otimes}V_{j} {\otimes}V_{j})/\SU(2)$. The basis states are tensor states, which we denote as
\begin{equation}
  |j, i_n\rangle = {\bigotimes}_{n}\ |i_n\rangle
\end{equation}
(by this we mean $|j, i_1, \ldots, i_5\rangle = |i_1\rangle \otimes \cdots \otimes |i_5\rangle$). We choose a basis in ${\cal I}_n$ as usual, that is: we fix a pairing of the links at each node and we choose the basis that diagonalises the modulus square of the sum of the $\SU(2)$ generators in the pair. 

In covariant LQG, the transition amplitude from an empty state to a state $|j,i_n\rangle$ in $\HH_j$ is given by the spinfoam amplitude of the boundary state $|j,i_n\rangle$ alone. The reason is that this transition corresponds to the amplitude of a boundary state that has \emph{only} one connected component, here interpreted as the future one. To first order in the spinfoam expansion, the amplitude of a boundary state is given by a single vertex. Hence the nothing-to-$|j,i_n\rangle$ amplitude is just the vertex amplitude for the boundary state
\begin{equation}
  \langle j, i_n|\varnothing\rangle= W(j,i_n) \equiv  \langle j, i_n|\psi_{o}\rangle
  \label{psizeo}
\end{equation}
where $W(j,i_n)$ is the Lorentzian EPRL vertex amplitude \cite{Engle2008}. 

Notice that this implies that we can view the ket $|\psi_{o}\rangle$ that has components $W(j,i_n)$ on the spin network basis, as the quantum state emerging from the big bang. This is the analogue, in (Lorentzian) LQG, of the Hartle-Hawking `no-boundary' initial state in (Euclidean) path integral quantum gravity \cite{Hartle1983}. This is the state we are interested in.  We want to study the mean geometry it defines and the quantum fluctuations and correlations it incorporates.  To this aim, we study the expectation value
\begin{equation}
   \langle A\rangle=\langle\psi_o|A|\psi_o\rangle,
\end{equation}
the spread
\begin{equation}
    \Delta A = \sqrt{\langle\psi_o|A^2|\psi_o\rangle-\langle A \rangle^2}
\end{equation}
and the (normalized) correlations
\begin{equation}
   C(A_1, A_2) = \frac{\langle\psi_o|A_1A_2|\psi_o\rangle-\langle A_1 \rangle \langle A_2 \rangle}{(\Delta A_1) \ (\Delta A_2)}
\end{equation}
of local geometry operators $A,A_1,A_2,...$ defined on $\HH$. When relevant, the dependence of the states and results from the scale parameter $j$ is denoted as $|\psi_o\rangle_j$ and $\langle A \rangle_j$. We compute also the entanglement entropy
\begin{equation}
  S = -\trace (\rho_n \log \rho_n)
\end{equation}
of a node with respect to the rest of the graph, where $\rho_n$ is the reduced density matrix of the state $|\psi_o\rangle$ at any node, all nodes being equivalent by symmetry.


\subsection{Quantum geometry}

The spin-network basis states $|j_l,i_n\rangle$ correspond to quantum tetrahedra (to quantum polyhedra in general \cite{Bianchi2011a} if the node have valence higher than 4). They can be viewed as a collection of quantum tetrahedra, glued identifying faces. Shared faces have the same area (but nor necessarily the same shape nor the same orientation of the faces). This geometry has been called \textit{twisted geometry} \cite{Freidel2010}. 

Consider an intertwiner state $|i_n\rangle$ of a 4-valent node $n$ with all spins equal to $j$ on the links entering the node. Let $\vec{E}_{nl}$ be one of the flux operators entering the node $n$ on link $l$
\begin{equation}
  \vec{E}_{nl} = (8\pi\,\gamma\,\hbar G) \vec{J}_l
\end{equation}
where $\gamma$ is the Barbero-Immirzi constant and $\vec{J}_l$ is the vector of $\SU(2)$ generators on link $l$. The flux operators entering the same node satisfies the Gauss constraint
\begin{equation}
  \sum_l \vec{E}_{nl} |i_n\rangle = 0.
\end{equation}
The areas of the faces are eigenvalues of the area operator
\begin{equation}
  A_{nl}|i_n\rangle = \sqrt{\vec{E}_{nl} \cdot \vec{E}_{nl}}|i_n\rangle = (8\pi\,\gamma\,\hbar G) \sqrt{j_l(j_l+1)}|i_n\rangle .
\end{equation}
The shape of the tetrahedron is measured by the angle operator
\begin{equation}
  \label{eq:geom-angleop}
  A_{ab} |i_n\rangle = \cos(\theta_{ab}) |i_n\rangle
\end{equation}
that gives the cosine of the (external) dihedral angle between faces $a$ and $b$. We obtain
\begin{equation}
  \label{eq:geom-anglejj}
  2 |\vec{J}_a| |\vec{J}_b| A_{ab} = 2 \vec{J}_a \cdot \vec{J}_b = (\vec{J}_a + \vec{J}_b)^2 - \vec{J}^2_a - \vec{J}^2_b.
\end{equation}
Suppose we have chosen the recoupling basis that pairs links $j_a$ and $j_b$ at node $n$, and let $|k_n\rangle$ denote the intertwiner state at node $n$ in such recoupling. The operator $(\vec{J}_a + \vec{J}_b)^2$ is diagonal on $|k_n\rangle$ with eigenvalue
\begin{equation}
  \label{eq:geom-casimir}
  (\vec{J}_a + \vec{J}_b)^2 |k_n\rangle = k_n(k_n+1) |k_n\rangle
\end{equation}
where $k_n$ is the intertwiner spin. Putting together eqs. \eqref{eq:geom-angleop}, \eqref{eq:geom-anglejj} and \eqref{eq:geom-casimir} we obtain
\begin{equation}
  \label{eq:geom-angleformula}
  \cos(\theta_{ab}) = \frac{k_n(k_n+1) - j_a(j_a+1) - j_b(j_b+1)}{2\sqrt{j_a(j_a+1)j_b(j_b+1)}}
\end{equation}
for measuring the dihedral angle $\cos(\theta_{ab})$ of $|k_n\rangle$ in terms of intertwiner spin $k_n$. Notice that since a spin-network state is the tensor product of intertwiner states there is no correlation between the different nodes and thus different shapes (angles) are not required to match. Moreover, according to the Heisenberg uncertainty principle, since different angle operators do not commute
\begin{equation}
  \label{eq:geom-commutator}
 [A_{ab}, A_{bc}] =  \frac{i\alpha}{2\,\vec{J}_a^2 \,|\vec{J}_b||\vec{J}_c|}\, \epsilon_{ijk} E^i_{na}E^j_{nb}E^k_{nc}
\end{equation}
then one of the dihedral angles will be sharp while the other angles will be spread. 

Using coherent states techniques \cite{Livine2007} it is possible to define a superposition of spin-network states that is peaked on a classical geometry with minimal spread. Other interesting states are maximally entangled states in which the entanglement between adjacent nodes forces gluing conditions on the corresponding quantum polyhedra \cite{Baytas2018}.  Here, instead, we are concerned with analysing the state $|\psi_o\rangle$, determined by the early universe dynamics in this truncation and defined in \eqref{psizeo}. This state can be seen as a quantum superposition of geometries.


\section{Amplitude}

The Lorentzian EPRL vertex amplitude $W(j_l, i_n)$ can be written as \cite{Speziale2017}
\begin{equation}
  W(j_l,i_n) = \sum_{l_f, k_e} \left(\prod_{e} \,(2k_{e}+1) 
    \,B(j_l, l_f; i_n, k_e) \right) \,\{15j\} (l_f, k_e)
\label{eq:vertexampl-full}
  \end{equation}
where $f$ and $e$ label the faces and the half-edges touching the vertex, the functions $B(j_l, l_f; i_n, k_e)$ are called \textit{booster functions} (see Appendix \ref{appendix:boost}) and $\{15j\}$ is the invariant $\SU(2)$ symbol built from contracting the five 4-valent intertwiners at the nodes (see Appendix \ref{appendix:wigner}). The product is over four of the five half-edges because one redundant factor must be eliminated by gauge-fixing. The sum is over a set of auxiliary spins $l_f$ and auxiliary intertwiners $k_e$. 

Analytical results show that in the large spin limit this amplitude is generally exponentially suppressed except in two cases \cite{Barrett2009, Dona2019}. The first case is when the boundary geometry is the geometry of the boundary of a Lorentzian 4-simplex.  This case can be naturally related to the semiclassical limit, where spacetime is flat and Lorentzian at scales smaller than the curvature radius.  The second case is when the boundary geometry is a vector geometry, which includes the case when the boundary geometry is the geometry of the boundary of a Euclidean 4-simplex.  As we see below, the mean geometry defined by $|\psi_o\rangle$ is that of a discretized metric 3-sphere, i.e. the boundary of a regular Euclidean 4-simplex. Therefore this is a vector geometry. Vector geometries have been considered as a puzzling feature of the theory \cite{Dona2018c}: here we see that they are a necessary contribution to the primordial quantum cosmological state in order to allow the tunneling from the empty state to the semiclassical 3-sphere geometry.


The form \eqref{eq:vertexampl-full} of the amplitude  is particularly suited for numerical evaluation on computers. The main difficulties come from: \begin{enumerate*}[label=(\roman*)]
  \item evaluation of the $15j$ symbol;
  \item evaluation of the booster functions;
  \item increasing computational complexity in the spin labels.
\end{enumerate*} We had to limit to maximum spins (scale factor) of the order $j \sim 15$ given the time and memory constraints imposed by our computing facility.

The sum over spins $l_f$ in eq. \eqref{eq:vertexampl-full} is unconstrained, as $l_f \geq j_l$, where link $l$ corresponds to face $f$. Hence it is necessary to introduce a cutoff $\Delta s$ so that
\begin{equation}
  l_f = j_l, j_l+1, \ldots, j_l + \Delta s
\end{equation}
and the exact value is found in the limit $\Delta s \to \infty$. The case with $\Delta s = 0$ has been called the \textit{simplified model} in \cite{Speziale2017}. Since the computation time is proportional to $(\Delta s+1)^6$, we had to limit ourselves to very low values of the cutoff. It can be shown that in the simplified model the Lorentzian part of the amplitude is partially suppressed \cite{Puchta2013,Speziale2017}, and the effect of increasing the cutoff $\Delta s$ is to gradually enhance the amplitude for Lorentzian configurations. For example, it appears that it is necessary to push the computations to higher cutoffs in order to match the expected behaviour in semiclassical asymptotics of Lorentzian simplexes \cite{Dona2019}. Nevertheless, we found that in our model, which spans the space of vector geometries due to the chosen (Euclidean) boundary conditions, the corrections due to higher cutoff values are minor or even negligible, so that the simplified computations effectively suffice to study the model numerically. All the computations were carried out using the \texttt{sl2cfoam} library \cite{Dona2018b}.


\section{Numerical results}

Here we report the results that we obtained numerically for the cosmological quantum state $|\psi_o\rangle$. The values are computed and shown for increasing values of the scale parameter $j$. When not stated otherwise, the Barbero-Immirzi constant $\gamma$ was set to the value $\gamma = 1.2$.

\subsection{Mean geometry}

We performed a numerical evaluation of the expectation value of the angle operator $A_{ab}$
\begin{equation}
  A_{ab} |i_n\rangle = \cos(\theta_{ab}) |i_n\rangle
\end{equation}
that measures the external dihedral angles between faces punctured by links $a$ and $b$ in any of the boundary tetrahedra (by symmetry all of them are equivalent). We found the expectation value
\begin{equation}
  \langle A_{ab}  \rangle = -0.333
\end{equation}
which corresponds to the cosine of the external dihedral angle of an equilateral tetrahedron, for any links $a,b$ chosen. This shows that the spatial metric of  $|\psi_o\rangle$ averages to that of the 3-boundary of a regular 4-simplex, i.e. to that of a 3-sphere in our approximation. The variation of the average with the scale parameter is minor and due entirely to numerical fluctuations. The independence of the result from the choice of the links was tested by switching to a different recoupling basis --- as the reducible basis of eq. \eqref{eq:15j-reducible} --- and also by directly performing the change of basis using eq. \eqref{eq:4jm-basis-change}.

\begin{figure}[h!]
  \raggedright
  \includegraphics[width=.9\columnwidth]{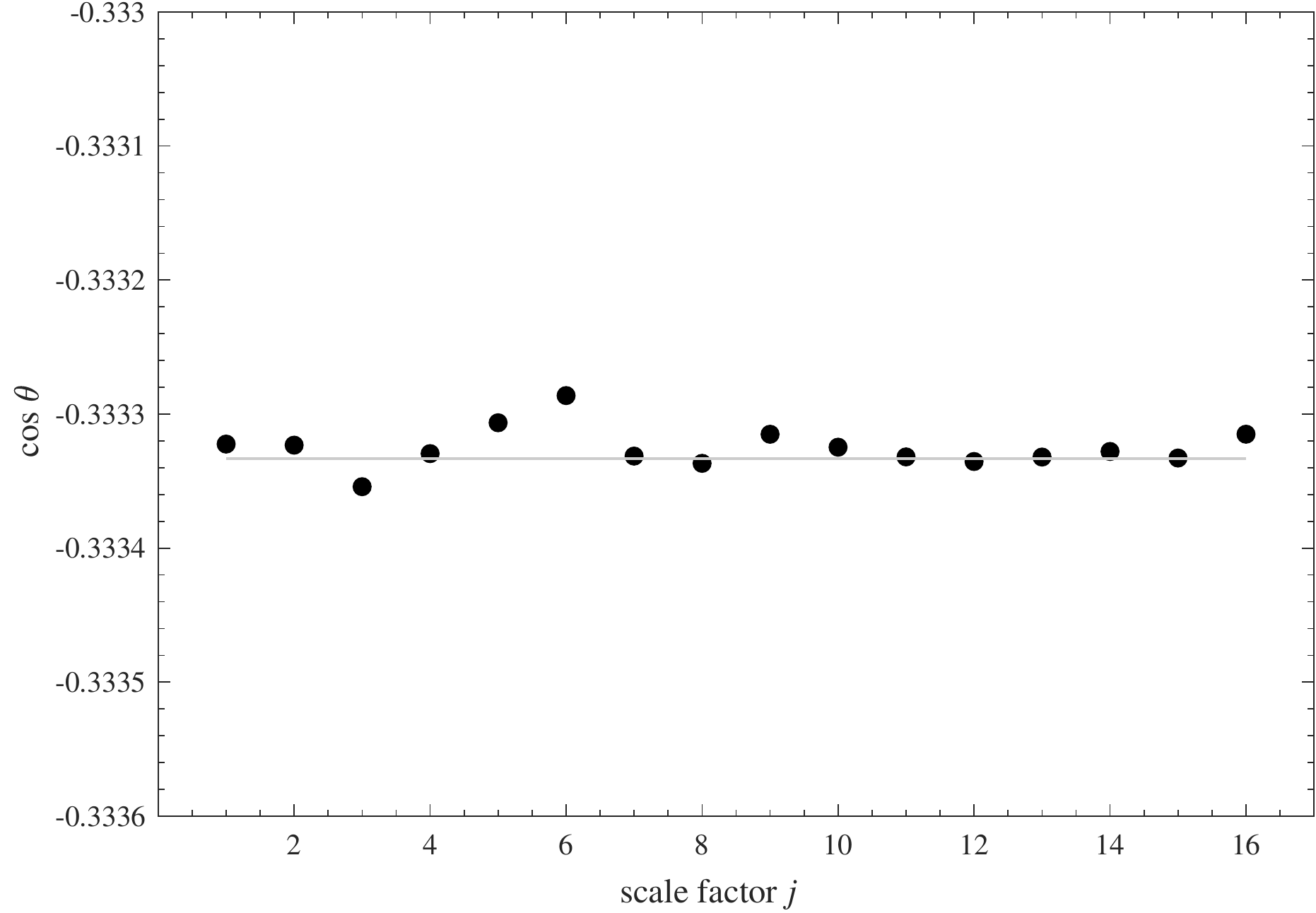} 
  \caption{The average external dihedral angle of boundary tetrahedra. The grey line shows the dihedral angle of a regular tetrahedron.}
  \label{plot:average-angle}
\end{figure}

\subsection{Fluctuations}

\begin{figure}[h]
  \hspace{-5mm}
  \includegraphics[width=.84\columnwidth]{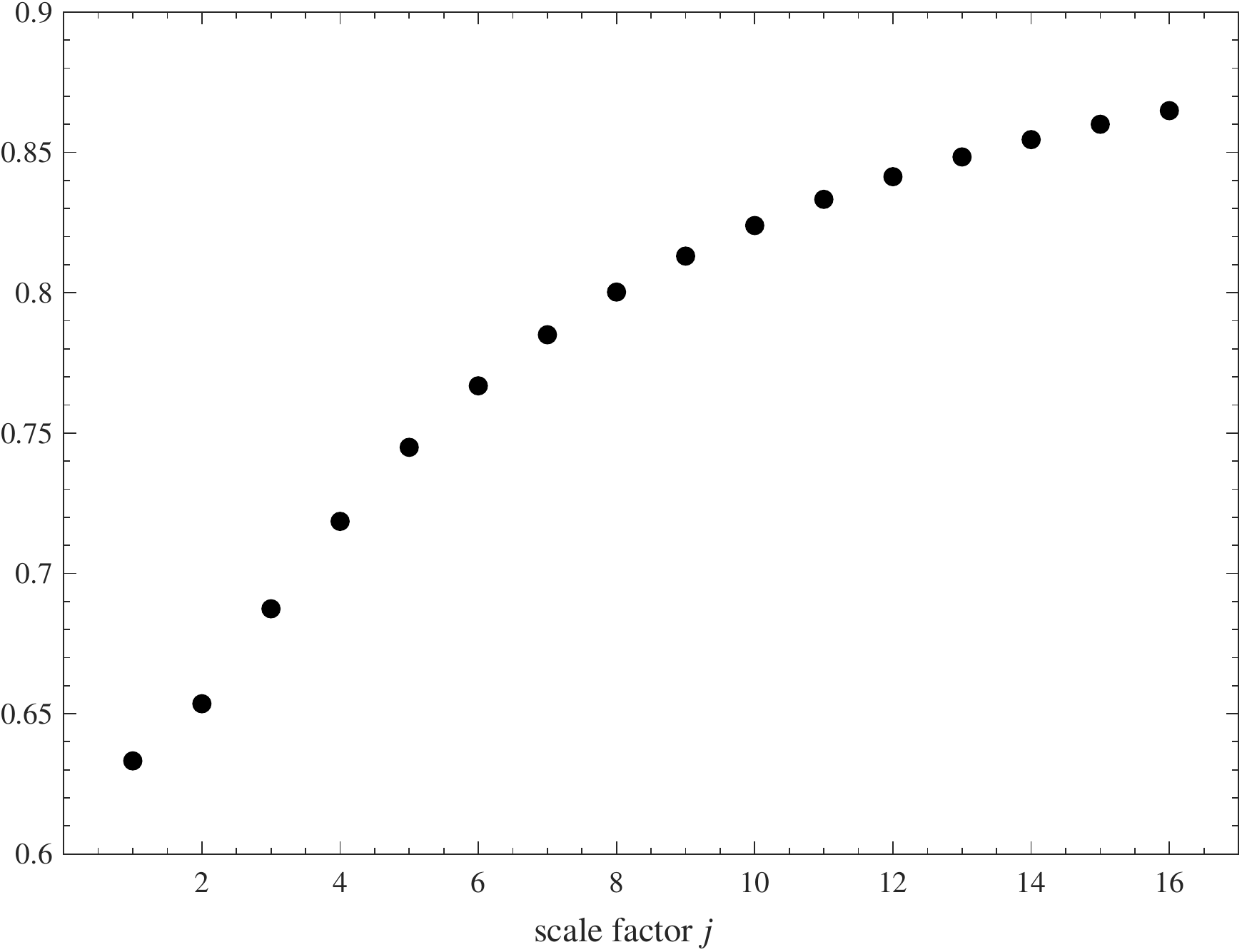} 
  \caption{Spread of angle operator.}
  \label{plot:spread-angle}
\end{figure}

The spread $\Delta A_{ab}$ is large and increasing with the scale factor, see Fig. \ref{plot:spread-angle}. This suggest that quantum fluctuations in the metric are wide and are not suppressed in the large-scale regime. These results are independent from the cutoff parameter $\Delta s$ and the value of the constant $\gamma$.
%


\subsection{Correlations and entanglement}

\begin{figure}[h!]
  \hspace{-4mm}
  \includegraphics[width=.84\columnwidth]{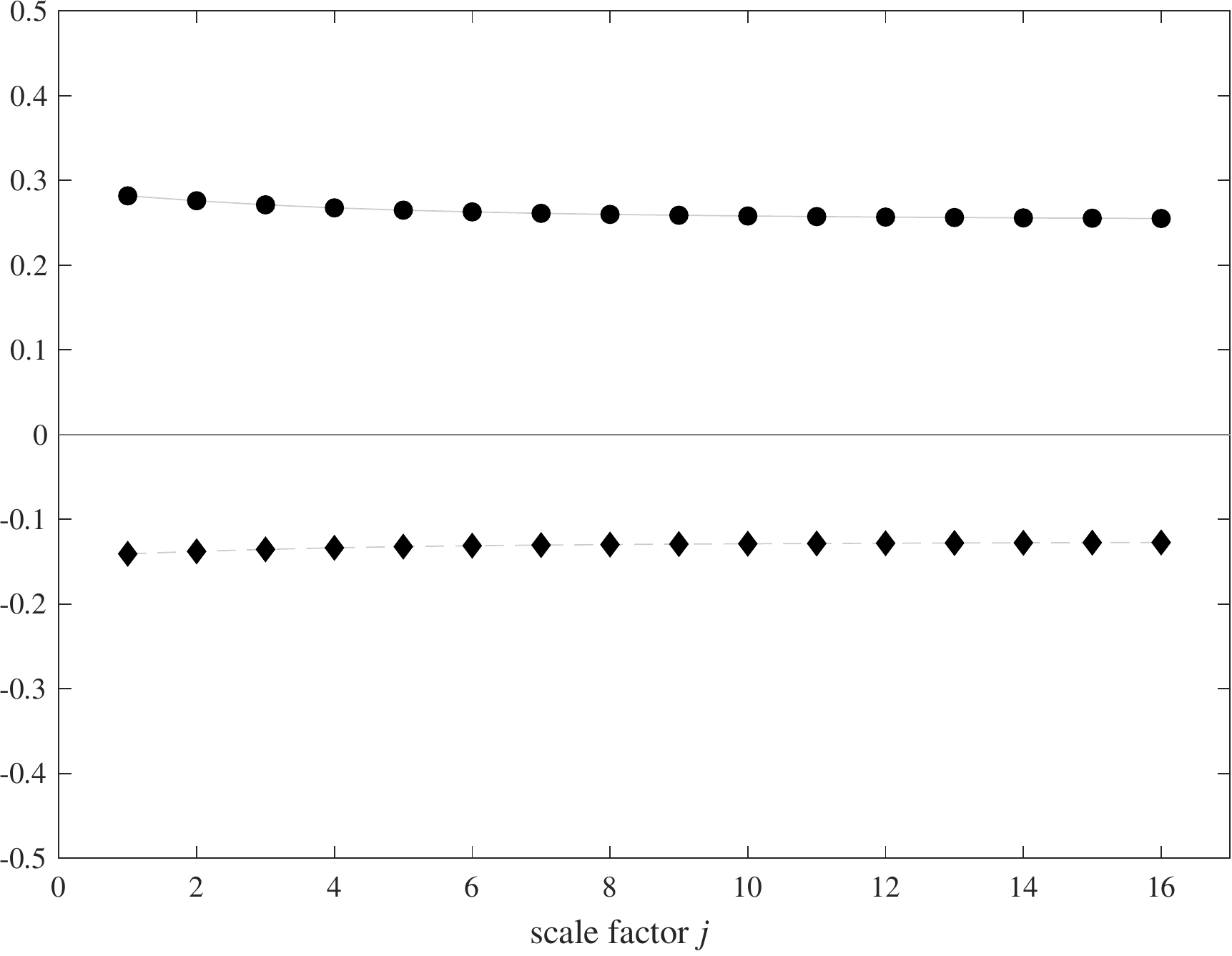} 
  \caption{Left: correlations of angle operator $A_{42,43}$ with $A_{23,24}$ (top, positive) and $A_{34,35}$ (bottom, negative). The same plot represents the correlations of $A_{41,45}$ with $A_{14,15}$ (top, positive) and $A_{54,53}$ (bottom, negative).}
  \label{plot:corr}
\end{figure}

We computed the correlations between angle operators on different nodes. The results now are not independent on the chosen links and nodes, because the chosen observables depend on the pairing. Let us introduce the notation
\begin{equation}
  A_{nn',nn''}
\end{equation}
for the angle operator $A_{ab}$ at node $n$, where now link $a$ connects $n$ with the node $n'$ and link $b$ connects $n$ with the node $n''$. Again, all source nodes $n$ are equivalent by symmetry (using the irreducible $15j$ symbol of eq. \eqref{eq:15j}). The correlations are shown in Fig. \ref{plot:corr}. Notice that for each source node $n$ there are thus two pairs of correlated and anti-correlated nodes. For example, for $n=4$ these are $(2,3)$ and $(1,5)$. What is more important is that the correlations (both positive and negative) seem to reach an asymptotic value, hence they are not suppressed in the large-scale regime. This means that the 3-metric that comes out from the quantum state $|\psi_o\rangle$ can correlate different spatial patches of the primordial universe, as required for solving the well-known horizon problem of standard cosmology.

\begin{figure}[h!]
  \hspace{-3mm}
  \includegraphics[width=.84\columnwidth]{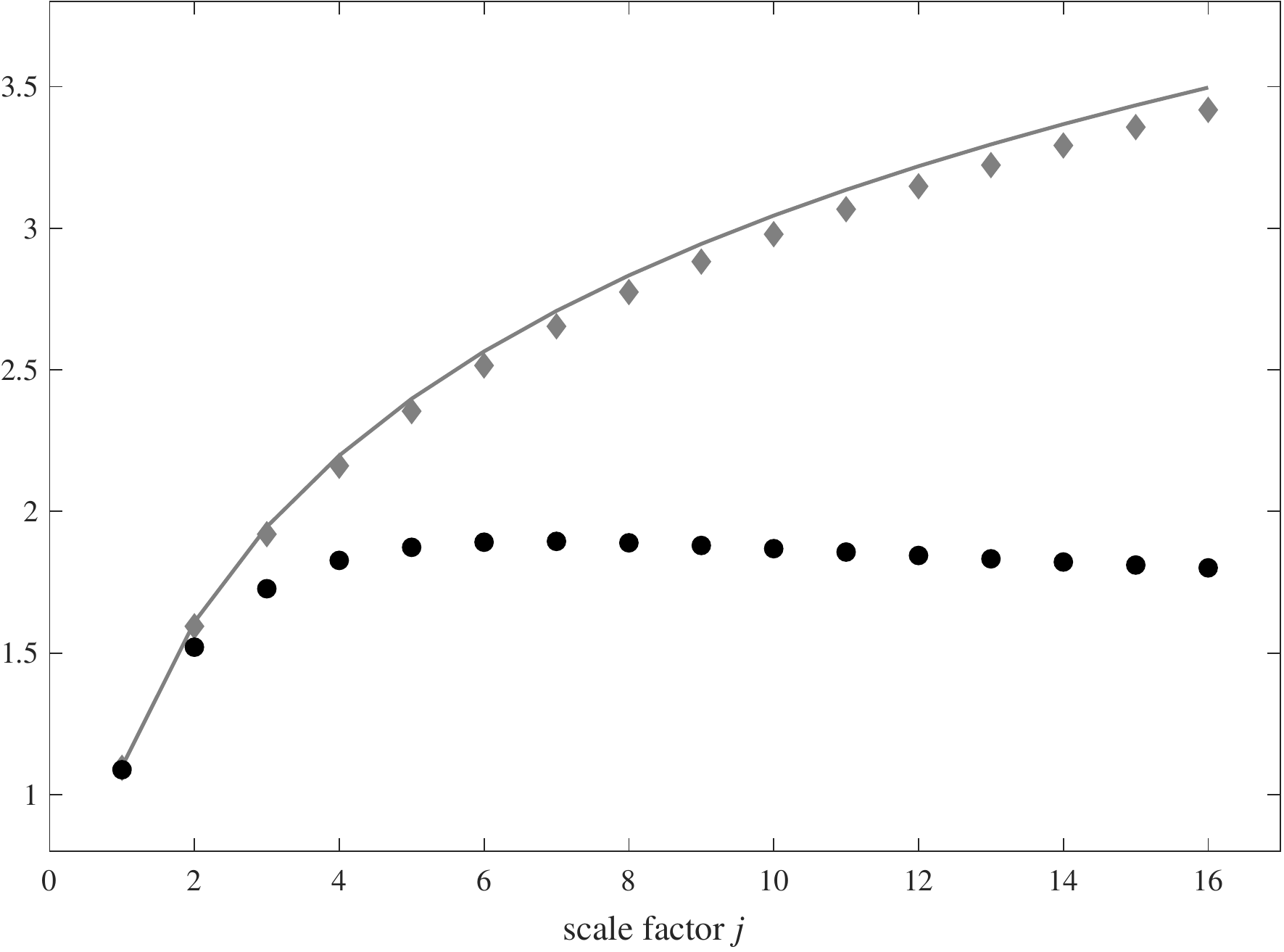} 
  \caption{The entanglement entropy of a node with respect to the rest of the graph. Grey continuous line shows the maximum entropy attainable for the given scale factor parameter. Grey diamonds  report the result of \cite{Bianchi2018}. Black circles show our result for $|\psi_o\rangle$.}
  \label{plot:ent}
\end{figure}

Another quantity to compute to quantify the degree of correlation between operators is the entanglement entropy between different nodes, viewed as quantum subsystems of the whole graph. A result by Page \cite{Page1993} states that, given a splitting $\HH = \HH_R \otimes \HH_{\bar{R}}$ of an Hilbert space $\HH$ into subspaces corresponding to a small subsystem $R$ and its complement $\bar{R}$, the typical state in $\HH_R$ is found to have an entanglement entropy equal to
\begin{equation}
  S_R \approx \log(\mathrm{dim} \HH_R)
\end{equation}
corresponding to a maximally-mixed pure state of the universe. In other words, the vast majority of quantum states of a small subsystem are close to being, in a broad sense, thermal (see also \cite{Popescu2006}). We can study the degree of non-tipicality of the primordial state $|\psi_o\rangle$ by looking at the entanglement entropy of any node as function of the scale factor. The result is shown in Fig. \ref{plot:ent}. The plot suggests that the entropy deviates significantly from the maximally-mixed case, and it appears to get close to an asymptotic value in the limit of large scale factor. Unfortunately, we could not push the computations to spins higher than $j = 16$, but the qualitative behaviour is clear. For comparison, we show also the maximum entropy
\begin{equation}
  S_\text{max}(j) = \log(2j+1)
\end{equation}
and the result of \cite{Bianchi2018} on the so-called \emph{Bell-network states} \cite{Baytas2018}, which are constructed in the same way as our primordial state $|\psi_o\rangle$ but using the dynamics of the much simpler \textit{BF theory}. It would be interesting to better understand the source of this highly non-trivial behaviour and to relate it to observable properties of our proposed primordial cosmological state.
%






\section{Conclusions}

We have studied the quantum state for the primordial universe predicted by the dynamics of Loop Quantum Gravity.  At first order in the LQG vertex expansion, the state describes the metric of a topologically closed, homogeneous and isotropic universe in its early quantum regime. Its degrees of freedom encode the shapes of neighbouring spatial regions. Their size (area) acts as a scale factor for our model and resembles the standard cosmological parameter $a(t)$. 

We studied the model numerically, in particular we studied the evolution of the operator which measures the angle between glued regions in the triangulated spacetime manifold. We studied the mean geometry as well as the quantum spread, the quantum correlations and the entanglement entropy of these regions. We found that the mean geometry is that of a 3-sphere, but the fluctuations are large, so that the geometry that emerges is only approximately regular. Moreover, we found that neighbouring regions are correlated and that these correlations do not vanish as the scale factor increases: this indicates that an inflationary phase may not be needed in order to circumvent the horizon problem, as the primordial quantum phase may introduce stochastic correlations in otherwise causally-independent spatial regions. In this respect, it would be enlightening to perform a finer analysis with a more complex triangulation in which long-range correlations may be distinguished from short-range correlations of neighbouring regions. 

Finally, we also computed the entanglement entropy of a single region viewed as a quantum subsystem of the whole universe. We found that our proposed cosmological state is highly non-typical, showing an entanglement entropy that is apparently reaching an asymptotic value as the scale factor increases. It would be interesting to link this peculiar behaviour to some geometrical properties of our model, and more ambitiously to some large-scale effect that could be tested by observations.

Our results indicate that an early quantum phase of the universe may provide a source and an explanation for known puzzling features of the standard cosmological model, such as the horizon problem, without introducing additional inflationary and/or bouncing phases. 

Our work is one of the first explorations of the purely quantum regime of LQG --- without resorting to study the high-spin semiclassical limit of the theory --- and also one of the first applications to a concrete physical model of the numerical tools that are recently being developed in the field of covariant Loop Quantum Gravity \cite{Bianchi2018,Dona2018b,Dona2019}.


\vskip15mm
\section{Acknowledgements}
The authors thank: Carlo Rovelli for extensive discussions and valuable remarks on this manuscript, Massimiliano Rinaldi and Simone Speziale for helpful insights on the earliest version of this work, and Giorgio Sarno for sharing his valuable knowledge on spinfoam numerics.
FG acknowledges support from INFN and thanks the Department of Theoretical Physics at UPV/EHU for the hospitality during his research internship. The research of FV at UPV/EHU was supported by the grant IT956-16 of the Basque Government and by the grant FIS2017-85076-P (MINECO/AEI/FEDER, UE). 
\vskip15mm


\appendix

\section{Recoupling coefficients}
\label{appendix:wigner}

We refer to the literature \cite{Varshalovich1988,Yutsis1962} for the definitions of the standard $3jm$, $6j$ and $9j$ symbols that appear in the recoupling theory of $\SU(2)$ representations. Here we provide the definition of the less common $4jm$ and $15j$ symbols that are used in our computations. 

Consider four $\SU(2)$ representations labelled by spins $j_1,j_2,j_3,j_4$. It is possible to combine the four spins in pairs in three possible ways, corresponding to three inequivalent recouplings. Each one of this recouplings corresponds to a different $4jm$-symbol for the given set of spins. For example, coupling spins $(j_1,j_2)$ and $(j_3,j_4)$ the definition of the symbol is
\begin{eqnarray}
  \wfj{j_1}{j_2}{j_3}{j_4}{m_1}{m_2}{m_3}{m_4}{j_{12}} &=&  \\ && \hspace{-3cm} \sum_{m_{12}}(-1)^{j_{12}-m_{12}} 
  \wtj{j_1}{j_2}{j_{12}}{m_1}{m_2}{m_{12}} \wtj{j_{12}}{j_3}{j_4}{-m_{12}}{m_3}{m_4}
  \nonumber
\end{eqnarray}
where the virtual spin $j_{12}$ is constrained by the inequality
\begin{equation}
  \max\left\lbrace |j_1-j_2|,|j_3-j_4| \right\rbrace \leq j_{12} \leq \min\left\lbrace j_1+j_2,j_3+j_4 \right\rbrace .
\end{equation}
A change in the recoupling basis involves summing over an intertwiner spin weighted by a $6j$ symbol. For example, the formula for a $4jm$ symbol in the recoupling basis that pairs $j_1$ with $j_3$ is
\begin{eqnarray}
  \label{eq:4jm-basis-change}
  \wfj{j_1}{j_3}{j_2}{j_4}{m_1}{m_3}{m_2}{m_4}{j_{13}} &=&  \\ && \hspace{-3cm} \sum_{j_{12}}\, (2j_{12}+1) \wsj{j_1}{j_3}{j_{12}}{j_4}{j_2}{j_{13}} \wfj{j_1}{j_2}{j_3}{j_4}{m_1}{m_2}{m_3}{m_4}{j_{12}}
  \nonumber
\end{eqnarray}
in terms of the $j_1$-with-$j_2$ basis. 

The $15j$ symbol is the contraction of five $4jm$ symbols over all independent magnetic indices. The choice of the recoupling basis determines if the symbol can be reduced to the product of lower-dimensional symbols. The most generic and the most symmetric symbol is the $15j$ symbol of \textit{first kind} which can be written as the contraction of five $6j$ symbols. In our convention the complete formula is
%
\begin{align}
  \label{eq:15j}
  \{15j\}&(l_f,k_e) = \sum_x \,(2x+1)\, (-1)^{\sum_a l_a + \sum_a k_a}\,  \\ \nonumber
 &\times \,\,  \wsj{k_1}{l_{25}}{x}{k_5}{l_{14}}{l_{15}} \, \wsj{l_{14}}{k_5}{x}{l_{35}}{k_4}{l_{45}} \wsj{k_4}{l_{35}}{x}{k_3}{l_{24}}{l_{34}} \\[5pt]
&\times \,\, \wsj{l_{24}}{k_3}{x}{l_{13}}{k_2}{l_{23}} \wsj{k_2}{l_{13}}{x}{k_1}{l_{25}}{l_{12}} \nonumber
\end{align}
where $l_{ab}$ denotes the spin on the link with source node $a$ and target node $b$. Whenever a symmetrical configuration of the angle operators was not required we also employed the following reducible symbol \cite{Dona2018c}
\begin{align}
  \label{eq:15j-reducible}
  \{15j\}_\text{Red}\,(l_f,k_e) &= (-1)^{2(l_{15} + l_{13} + k_1) - (k_2 + k_3 + k_4 + k_5)}  \nonumber \\
  &\times\,\wsj{k_1}{k_3}{k_2}{l_{23}}{l_{12}}{l_{13}}\wsj{k_1}{k_4}{k_5}{l_{45}}{l_{15}}{l_{14}}
  \nonumber \\
  &\times\,\wnj{k_2}{k_3}{k_1}{l_{24}}{l_{34}}{l_{15}}{l_{25}}{l_{35}}{k_5}
\end{align}
whose evaluation on our hardware was roughly one order of magnitude faster than the irreducible symbol. The numerical computations of the $3jm$, $6j$ and $9j$ symbols were performed using the \texttt{wigxjpf} library \cite{Johansson2015}.

\section{Booster functions}
\label{appendix:boost}

The booster functions $B(j_l, l_f; i_n, k_e)$ were initially studied in the context of representation theory of the Lorentz group $\SL(2,\CC)$ \cite{Ruhl1970}. In their most general formulation, they are the $\SL(2,\CC)$ analogues of the usual Clebsch-Gordan coefficients for the rotation group $\SU(2)$. The EPRL-FK model of LQG fixes some of the free parameters of this general functions via the so-called \textit{Y-map}, which is induced on the infinite-dimensional irreducible representation of $\SL(2,\CC)$ by the \textit{simplicity constraint} in the Holst formulation of general relativity \cite{Rovelli2015a}. The formula for the booster functions (on a 4-valent strand) then reads
\begin{eqnarray}
\label{eq:booster}
B(j_l, l_f; i_n, k_e) &=& \sum_{p_i}
 \left(\begin{array}{c} j_l \\  p_i \end{array}\right)^{(i_n)} \left(\begin{array}{c} l_f \\ p_i \end{array}\right)^{(k_e)}  \\ \nonumber &&\times \left(\int_0^\infty \mathrm{d} r \;  \frac{\sinh^2r}{4\pi}\ \prod_{i=1}^4 \ d^{(\gamma j_i,j_i)}_{j_il_ip_i}(r)\right)
\end{eqnarray}
which is formed by two $4jm$-symbols (see Appendix \ref{appendix:wigner}) and an integration of the boost matrix elements $d^{(\gamma j_i,j_i)}_{j_il_ip_i}(r)$. The boost matrix element in our phase convention \cite{Dona2018b,Speziale2017} is
\begin{align}
\label{eq:d-small}
&d^{(\gamma j,j)}_{jlp}(r) =  (-1)^{\frac{j-l}2} \\ \nonumber &\times \frac{\Gamma(j+i\gamma j+1)}{|{\Gamma(j+i\gamma j+1)}|} \frac{\Gamma(l-i\gamma j+1)}{|{\Gamma(l-i\gamma j+1)}|} \frac{\sqrt{2j+1}\sqrt{2l+1}}{(j+l+1)!}  \\  
&\times e^{-(k-i\gamma j +p+1)r} \left[ (2j)!(l+j)!(l-j)!\frac{(l+p)!(l-p)!}{(j+p)!(j-p)!}\right]^{1/2}  \nonumber \\ 
&\times \sum_{s} \frac{ (-1)^{s} e^{- 2 sr} }{s! (l-j-s)!}\,  
{}_2F_1\!\left[l\!+\!1\!-\!i\gamma j,j\!+\!p\!+\!1\!+\!s,j\!+\!l\!+\!2,1\!-\!e^{-2r}\right] 
\nonumber
\end{align}
where $\,{}_2F_1[a, b, c, z]$ is the hypergeometric function. It is possible to rewrite the formulae \eqref{eq:booster} and \eqref{eq:d-small} into a form that is more tailored for the numerical evaluation on computers. In particular, a formula which trades the integration over $r$ with a simpler integration of a smooth function of a virtual irrep label has been proposed in \cite{Speziale2017} and tested on our model for low spins. The \texttt{sl2cfoam} library \cite{Dona2018b} uses a different formula which was proposed in \cite{Collet}.


\medskip

\providecommand{\href}[2]{#2}\begingroup\raggedright\endgroup

\noindent
$\ast$\! \small \it Disclaimer: This bibliography has no presumption of completeness. Suggestions for improvements are welcome. 

\end{document}